\documentclass[preprint]{aastex}

\newcommand{\cps}{${\rm count\,s}^{-1}$}

\shortauthors{Galloway}
\shorttitle{Accretion column disruption in GX~1+4}

\begin{document}


\title{Accretion column disruption in GX~1+4}


\author{D.~K. Galloway\altaffilmark{1}}
\affil{School of Mathematics and Physics, University of Tasmania, Hobart 7001,
Australia}
\affil{RCfTA, School of Physics, University of Sydney, NSW 2006 Australia}


\altaffiltext{1}{present address: Center for Space Research, MIT,
    37-571, 77 Massachusetts Avenue, Cambridge, MA 02139-4307}


\begin{abstract}
Daily observations of the binary X-ray pulsar GX~1+4 were made with the
{\it Rossi X-ray Timing Explorer\/} ({\it RXTE\/}) satellite between 1997
May 16--20 as part of a four-month monitoring program.  On May 17 the
sharp dips normally observed in the lightcurve were all but absent,
resulting in a pulse fraction $f_{\rm p}\approx 0.5$ instead of the more
typical value of $\approx 0.8$ measured before and after. Also observed
was a dramatic hardening of the 2--40~keV phase-averaged spectrum. The
power-law photon index was $1.16\pm0.02$, whereas values of 1.6--2.0 are
more typical. In terms of a Comptonization continuum component, the
optical depth for scattering was $\tau\approx 19$, with 4--6 the usual
range for {\it RXTE\/} spectra \cite[]{myphd:a}.  Pulse-phase spectrosopy
indicates that $\tau$ is decreased relative to the phase-averaged value
around the primary minimum, where an increase is normally observed.  The
reduced depth of the dip is interpreted as disruption of the accretion
column, and the accompanying spectral variation suggests a substantially
different accretion regime than is usual for this source.
\end{abstract}




\section{Introduction}

The X-ray pulsar GX~1+4 is unusual in several respects. It is the only
confirmed neutron star in a symbiotic binary \cite[]{symcat00} which
appears to be accreting from the giant companions' stellar wind. The
neutron star spin period evolution exhibits long periods of almost
constant spin-up or -down, punctuated by `torque reversals' lasting tens
of days \cite[e.g.][]{nel97}.  Indirect estimates indicate a surface
magnetic field strength of 2--3$\times 10^{13}$~G
\cite[]{beu84,dot89,mony91,gre93,cui97}.  These values are significantly
greater than the $\sim10^{12}$~G inferred for other accreting X-ray
pulsars, generally from measurements of cyclotron scattering resonance
features \cite[CSRFs; e.g.][]{heindl00} in the X-ray spectra.  As the
presence of such a strong field in GX~1+4 awaits confirmation by such
methods, the possibility that the field strength may in fact be one or
more orders of magnitude lower remains.

Historically X-ray pulsar spectra have generally been fitted with
variations on a power law, which is well known to be a signature of
unsaturated Comptonization \cite[e.g][]{pss83}.  Model fitting to the
available archival {\it RXTE\/} data show that the spectra of GX~1+4 are
broadly consistent with an analytic approximation to a Comptonization
continuum component combined with a gaussian to represent iron line
emission, both attenuated by neutral absorbtion with variable column
density \cite[]{myphd:a}.

The archival {\it RXTE\/} data, which span two orders of magnitude in flux
for the source, has also revealed that the X-ray lightcurves and mean
pulse profiles of all but one of the observations exhibit sharp dips
spanning $\Delta\phi\approx0.1$--0.2 in phase.  During 1996 July the
countrate dropped to $\approx 10$~\cps\ over $\approx6$~h; sharp dips were
observed in the mean pulse profiles even then \cite[]{gil00}.  Similar
dips are observed in only two other X-ray pulsars, RX~J0812.4-3114
\cite[]{rei99} and A~0535+262 \cite[]{cem98}.  A recent analytic and
modelling study suggests that the sharp dips are signatures of eclipses of
the emitting region by the accretion column (Galloway et al. 2000, in
preparation).  Pulse-phase spectroscopy using a continuum spectral model
simulating thermal Comptonization of both GX~1+4 and RX~J0812.4$-$3114
shows a significant increase in the fitted optical depth $\tau$ coincident
with the dip phase.  If the inclination angle $i$ of these systems are
roughly equal to their magnetic colatitude $\beta$, such an increase may
result when one of the accretion columns approaches alignment with the
line of sight once each rotation period.  At the phase of closest
alignment, photons emitted from the polar cap must propagate a greater
distance through the column to escape and ultimately reach the observer.
Numerical modeling using a simplified column geometry confirms that the
resulting increase in optical depth will have just the effect observed in
GX~1+4 and RX~J0812.4$-$3114 \cite[]{gal00:beam}.


\section{Observations and Data Analysis}

{\em RXTE\/} consists of three instruments, the proportional counter array
(PCA) sensitive to photons in the energy range 2--60~keV, the high-energy
X-ray timing experiment (HEXTE) covering 16--250~keV, and the all-sky
monitor (ASM) which spans 2--10~keV \cite[]{giles95}.
Analysis of {\it RXTE\/} data presented in this letter was carried out using
{\sc lheasoft~5.0}, released 23 February 2000 by the {\it RXTE\/} Guest
Observer Facility (GOF). The data were first screened to ensure that the
pointing offset was $<0.02\arcdeg$ and the source was $>10\arcdeg$ from the
sun. This introduces additional gaps to the data.
Instrumental background from cosmic ray interactions and as a result of
satellite passages close to the SAA are estimated using the {\sc pcabackest}
software which is included in the {\sc lheasoft} package.  The `bright'
source background models from the 23 February 2000 release were used,
which are appropriate when the net source countrate is
$\ga 60$~\cps.  Spectral fitting was undertaken using
the {\sc xspec} spectral fitting package version 11 \cite[]{xspec}.
For more details of the spectral analysis see \cite{myphd:a}.

\label{P20170}

\begin{figure}
 \epsscale{0.9}
 \plotone{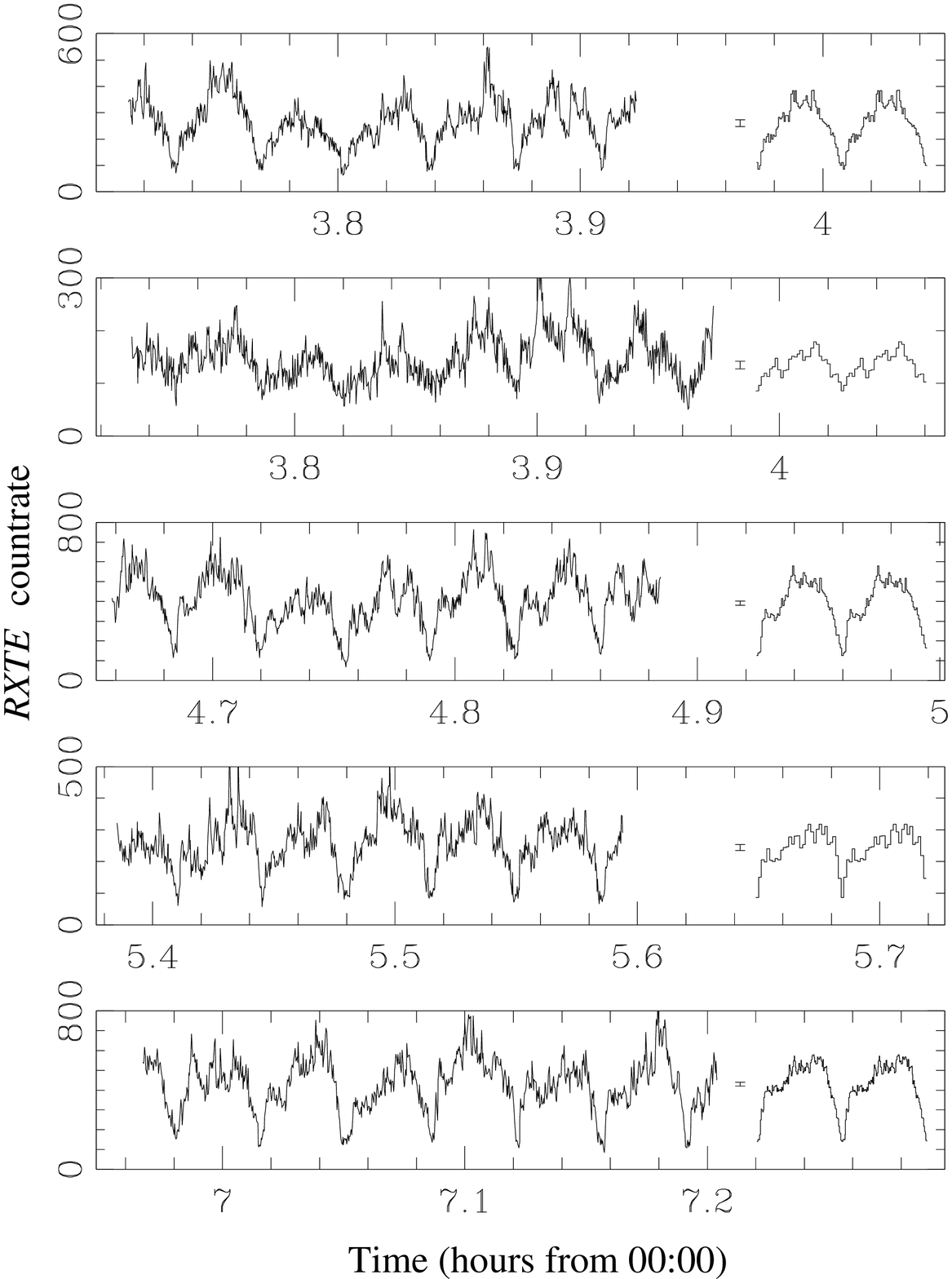}
\figcaption[p20170-Y-new.eps]{ {\it RXTE\/} observations of GX~1+4, 1997
May 16--20. Each panel shows the countrate on 1~s bins for all 5 PCUs for
the short observations on May 16, 17, 18, 19 and 20 from top to bottom
respectively. Shown at the right are the corresponding mean pulse profiles
(over two full cycles) with a representative error bar. The profiles are
calculated by folding the 1~s lightcurves on the pulse period $126.4319(0)
\pm 0.0005(5)$~s estimated from regular BATSE monitoring of the source.
Note the change in scale on the $y$-axis between the panels.
\label{P20170-Y}}
\end{figure}

Between 1997 May 16--20 GX~1+4 was observed daily for $\approx 12$~min.
The phase-averaged background subtracted countrate varied from day to day,
from $\la 100$~\cps\ (with all 5 proportional counter units operating) on
17 May to $\approx 500$~ \cps\ on 18 and 20 May (Figure \ref{P20170-Y}).
For four of the five observations the sharp dips which form the primary
minimum in the pulse profile were clearly observed in the background
subtracted PCA countrate. The pulse fraction is defined as 
\begin{eqnarray}
  f_{\rm p} = \frac{F_{\rm max}-F_{\rm min}}{F_{\rm max}}
	\hspace{1cm}{\rm where} \nonumber \\
  F_{\rm min} < F(\phi_i) < F_{\rm max}\ \ \forall\ i\in{1\ldots n} \nonumber
\end{eqnarray}
and where $F(\phi_i)$ is the mean pulse profile calculated over $n$ bins.
On May 16 and 18--20 $f_{\rm p}\approx0.8$, as is typical for the source.
On 17 May however, the sharp dips in the pulse profile were weaker or
absent, resulting in a pulse fraction $f_{\rm p}\approx 0.5$ (Figure
\ref{P20170-Y}, second panel from top).  The minima are instead broad and
irregular, lacking the consistency over successive cyles observed on other
days.


Spectral model fitting of the Standard-2 mode spectra averaged over each
daily observation was undertaken with the Comptonization continuum model
\cite[`{\tt compTT}' in {\sc xspec};][]{tit94} following \cite{myphd:a}.
Note that it was not possible to obtain a background spectrum for the May
20 observation due to problems with the filter file; the background
spectrum from the previous days' observation was used instead.  Several of
the spectral fit parameters show significant variability (Table
\ref{tab-P20170}).  The column density $n_{\rm H}$ varied between $\approx
10^{23}\ {\rm cm}^{-2}$ and $\approx 3.2\times 10^{23}\ {\rm cm}^{-2}$.
The source spectrum temperature $T_0$ was somewhat higher than is typical
on 16 May but decreased to a more typical 1.2--1.3~keV by 18 May.  For two
of the spectra the scattering plasma temperature $T_{\rm e}$ similarly
could not be constrained by the fitting algorithm, and instead was fixed
at 10~keV.  For the other three days $T_{\rm e}$ varied between $\approx
6.5$--7 and $\approx 10$~keV.  The most dramatic variation was in the
fitted optical depth $\tau$, which was 3.6--4 on May 16 and 18--20 but
$\approx 19$ on May 17. The equivalent width of the Fe gaussian component
was also unusually high for this spectrum, at more than 1.8~keV.  These
variations occurred as the luminosity varied between 1.5--$3.6\times
10^{37}\ {\rm erg \,s}^{-1}$, assuming a source distance of 10~kpc. The
actual distance of the source is thought to be in the range 3--15~kpc
\cite[]{chak97:opt}.

The $\chi^2_\nu$ values indicate good fits for each spectrum except on May
17.  The spectrum obtained on this day does not resemble any of the other
$\approx 50$ spectra extracted over the course of the {\it RXTE\/}
observations between 1996 and 1997.  The model fit was problematic in
several respects.  Initial fits with all parameters free to vary resulted
in a fit value for the $\tau_{\rm max}$ parameter (associated with the Xe
absorption edge at 4.83~keV) an order of magnitude greater than the
maximum for all the other spectra. Thus the value for this parameter was
instead fixed (`frozen') at the fit value for the spectrum extracted over
the whole of the 1997 May observation.  The fitted $T_0$ value, when free
to vary, also became much larger than was consistent than for the other
spectra and so was fixed at 1.3~keV. This value is typical for the source
at other times \cite[]{myphd:a}.  While the  resulting fit statistic
$\chi^2_{\nu}\approx2.60$ indicates an unacceptable fit, no improvement
was found using alternative continuum models, including power law, power
law with exponential cutoff and broken powerlaw.  A comparable fit was
obtained with a blackbody component resulting in $\chi^2_\nu=2.74$ with
fitted $T_{\rm bb}=6.65^{6.76}_{6.54}$~keV in rough agreement with $T_{\rm
e}$ for the {\tt compTT} fit of $6.51^{6.60}_{6.40}$~keV.

\begin{figure}
 \epsscale{0.9}
 \plotone{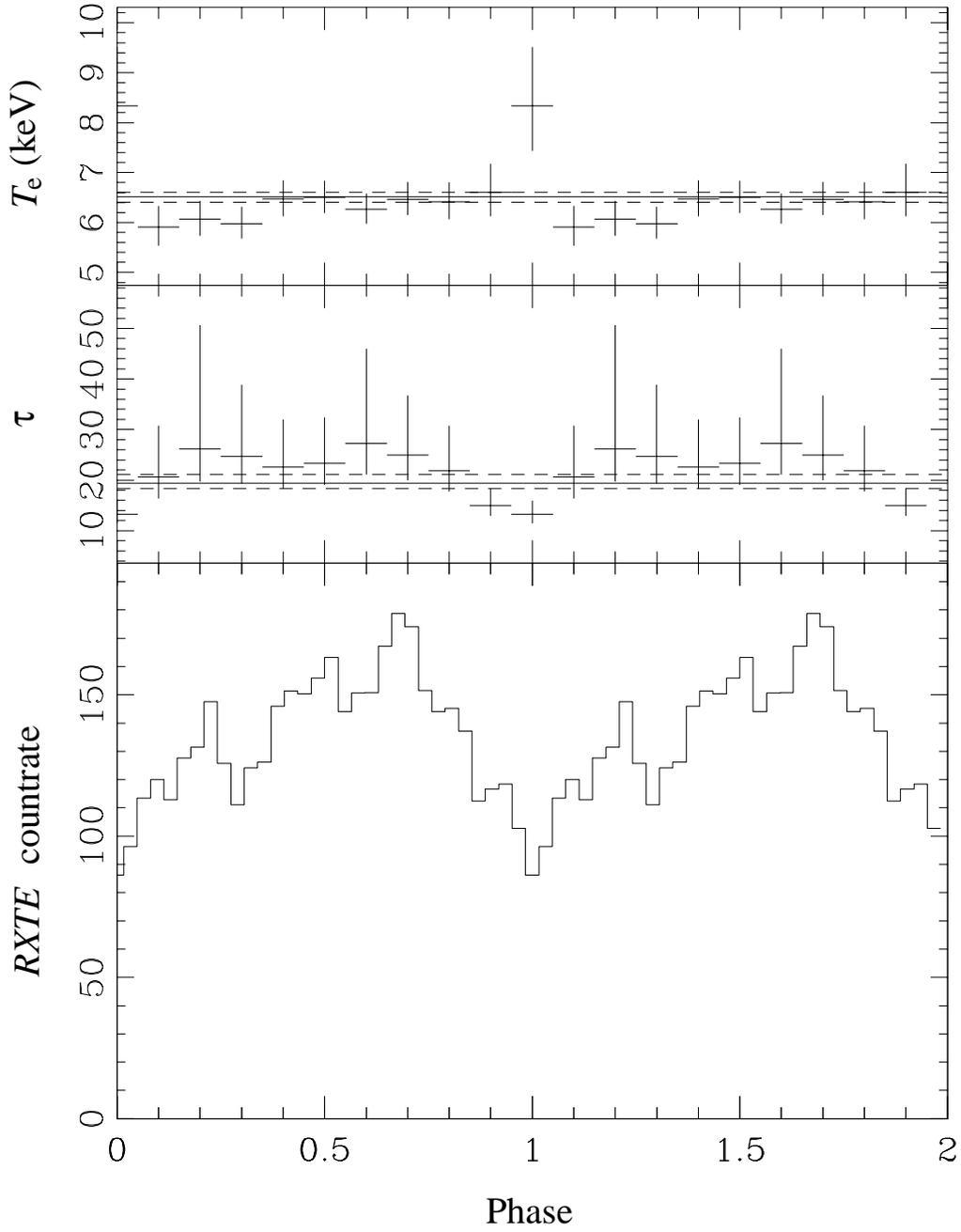}
\figcaption[pps-Y2.eps]{ Pulse-phase spectroscopy of data from the {\it
RXTE\/} observation of GX~1+4 on 1997 May 17.  The top and middle panels
show the variation in fitted $T_{\rm e}$ and $\tau$ respectively with
phase.  The solid and dotted lines show the fitted values and 90~per~cent
uncertainty limits for the phase-averaged spectrum.  Errorbars represent
90~per~cent uncertainty limits.  The bottom panel shows the pulse profile
over the full PCA energy range (as shown in Figure \ref{P20170-Y}).
\label{pps-Y2}}
\end{figure}

The data from each of the observations on May 16--20 were divided into 10
phase bins based on the pulse period $P= 126.4319(0) \pm 0.0005(5)$~s
estimated from regular BATSE monitoring of the source.  The ephemeris for
each day was chosen arbitrarily so that the first phase bin was centered
on the primary minimum, defined as phase 0.0.  Each of the resulting
phase-selected spectra were fitted with the usual spectral model with at
most $T_{\rm e}$, $\tau$ and the {\tt compTT} and gaussian component
normalisations free to vary.  The other parameters were fixed at the fit
values for the phase-averaged spectrum on the same day.  This approach is
essentially identical to that of \cite{gal00:spec}.  For the May 16 and 19
observations $T_{\rm e}$ was also fixed at the same value as for the
phase-averaged spectrum (see Table \ref{tab-P20170}).  Variation of
$T_{\rm e}$ and $\tau$ with phase is small except close to the primary
minimum. The fitted $\tau$ around phase 0.0 on May 16 and 18--20 was
significantly greater than the value for the phase-averaged spectrum, by
20--60~per~cent. For the May 17 observation, at the much shallower primary
minimum $\tau$ is instead marginally lower than the phase-averaged fit
values, while $T_{\rm e}$ is somewhat higher (Figure \ref{pps-Y2}).

\section{Discussion}

For sources with $i\approx\beta$, assuming that $i$, $\beta$ and the
footpoints of the accretion column on the neutron star do not change, dips
due to column eclipses must always be observed while accretion continues
through the columns.  The apparent disappearance of the dip during the
1997 May 17 observations clearly challenges this picture. The flux at this
time is more than an order of magnitude larger than the lowest level
measured by {\it RXTE\/}; complete cessation of accretion seems unlikely.
Another possibility is that the geometry of the system is changing such
that the accretion column drops out of alignment briefly.  The inclination
angle $i$ may vary due to geodetic precession of the pulsar, as is
suggested for some radio pulsars \cite[e.g.][]{kramer98}.  Alternatively,
the magnetic colatitude $\beta$ may change due to evolution of the neutron
star magnetic field.  Variations due to the former effect, however, are
only expected on timescales of years. And while the latter possibility
cannot be discounted due to the lack of knowledge regarding such
evolution, it seems much more likely that each of these mechanisms would
cause a much longer-lasting, if not permanent, cessation of eclipses
rather than what is observed in GX~1+4.

Variations in the position of the column itself are much more plausible.
As the accretion rate $\dot{M}$ (and hence the luminosity $L_{\rm X}$)
decreases, the balance of magnetic and gas pressures at the inner edge of
the accretion disc means that the disk will be truncated further out from
the neutron star \cite[]{gl79a,gl79b}. Field lines threading the boundary
region where disk material decelerates and becomes bound to the accretion
column will thus meet the neutron star at progressively higher latitudes
on the star. In principle at least this mechanism may cause the column to
move in and out of alignment, in which case eclipses (and dips) would be
expected only over a finite range of $L_{\rm X}$. Instead what appears to
be the case for GX~1+4 is that dips are normally observed at the full
range of luminosities at which the source was observed by {\it RXTE}, and
are perhaps absent at a particular flux level only within that range.
Observations at flux levels comparable to that of 1997 May 17 however do
generally exhibit sharp dips. If the accretion column has indeed been
disrupted rather than moved, accretion must be occurring over a much
larger region of the neutron stars' magnetosphere to give the observed
X-ray flux.  That material can enter the magnetosphere of the star without
being channeled along the field lines suggests either that the magnetic
field strength is in actual fact low for an X-ray pulsar, or that some
global magnetohydrodynamic instability is operating which can completely
disrupt the flow.

The unique phase-averaged {\it RXTE\/} spectrum observed coincident with
the unusually low pulse fraction underlines the significance of this
event. That the optical depth has increased fivefold compared to values
obtained before and after suggests that the source has been `smothered' by
hot accreting material, which presumably also contributes to the unusually
strong Fe line emission.  Such a large $\tau$ suggests that the photon
spectrum will approach thermalization with the scattering material; the
close agreement of the fitted blackbody temperature and the scattering
plasma temperature provides corroboration for the {\tt compTT} component
fits during this interval. The pulse-phase spectral variation was also
different to that normally observed in the source. Typically $\tau$ is
significantly greater than the mean value around the primary minimum; this
is the case for previously published results \cite[e.g.][]{gal00:spec} as
well as the observations on May 16 and 18--20. On May 17 $\tau$ was
instead somewhat {\em less} than the phase-averaged value around the
primary minimum. These results are strongly suggestive of a substantially
different distribution of matter around the source.  The increase in the
equivalent width also observed on May 17 was not accompanied by the usual
increase in the neutral column density $n_{\rm H}$ \cite[]{kot99,myphd:a},
further indicating varying conditions close to the star rather than in the
outer circumstellar matter.

At least one similar event has been observed previously from this source.
On 1993 December 11 a pulse fraction $\approx 35$~per cent (20--100~keV)
was measured using balloon-borne instruments, along with an unusually hard
spectrum with power law energy index $0.54\pm0.18$ \cite[]{rao94}.  For
comparison, a power law fit was made to {\it RXTE\/} spectra from 17 May
in the energy band 20--40~keV only. The resulting photon index
$1.15_{1.03}^{1.28}$ is consistent with the fit value for the full PCA
energy range, and represents a significantly harder spectrum than on 1993
December 11.  Flux measurements by the Burst and Transient Source
Experiment (BATSE) aboard the {\it Compton Gamma-ray Observatory}
satellite indicate that following that observation the source entered a
hard X-ray low state with $L_{\rm X}\la0.1\ {\rm keV\,cm^{-2}\,s}^{-1}$
(20--60~keV) which persisted (with brief interruptions) for the next
$\approx250$~d. In contrast, the May 17 observation was made during a
flare lasting $\approx20$~d where the flux peaked at $0.2\ {\rm
keV\,cm^{-2}\,s}^{-1}$.  The other spectra from 1997 May give fitted
$\tau$ values at intermediate levels between the two distinct spectral
states suggested by analysis of the archival {\it RXTE\/} observations
\cite[]{myphd:a}.  It is possible that the disruption of the accretion
column is also related to transitions between distinct spectral and
luminosity states inferred from analysis of the full set of {\it RXTE\/}
archival data \cite[]{myphd:a}.

\acknowledgments

This research has made use of data obtained through the
High Energy Astrophysics Science Archive Research Center Online Service,
provided by the NASA/Goddard Space Flight Center, and also the BATSE pulsar
group home page at \url{http://www.batse.msfc.masa.gov}.

\clearpage



\clearpage

\begin{deluxetable}{lccccc}
\tablecaption{Spectral fit parameters using the Comptonization continuum
component for the May 1997 {\it RXTE\/} observations of GX~1+4.
Fit parameters are the absorption column density $n_{\rm H}$ (in units of
$10^{22}\ {\rm cm}^{-2}$), source spectrum temperature $T_0$, 
plasma temperature $T_{\rm e}$ and optical depth $\tau$, Comptonization
component normalisation $A_{\rm C}$ (in units of $10^{-2}\ {\rm
photons\,cm^{-2}\,s^{-1} \,keV^{-1}}$), gaussian component center energy
$E_{\rm Fe}$, width $\sigma$, normalisation $A_{\rm Fe}$ (in units of
$10^{-3}\ {\rm photons\,cm^{-2}\,s^{-1}\,keV^{-1}}$) and equivalent width
EW, and finally the 2--60~keV flux $L_{\rm X}$ in units of $10^{37}\ {\rm
erg \,s}^{-1}$ (assuming a distance of 10~kpc). Energy range for fitting is
typically 2.8--35~keV.
    \label{tab-P20170} }
\tablewidth{0pt}
\tablehead{
 \colhead{Parameter} & \colhead{May 16} & \colhead{May 17} & \colhead{May 18}
  & \colhead{May 19} & \colhead{May 20} } 
\startdata
$n_{\rm H}$ 
   & $17.7^{21.0}_{14.9}$ & $10.6^{11.9}_{9.33}$ & $21.4^{24.7}_{17.8}$
   & $31.9^{35.5}_{26.2}$ & $21.1^{24.2}_{20.0}$  \\
$T_0$ (keV)
   & $1.515^{1.662}_{1.351}$ & 1.3 (fixed) & $1.272^{1.424}_{1.139}$
   & $1.231^{1.498}_{1.073}$ & $1.292^{1.378}_{1.171}$  \\
$T_{\rm e}$ (keV)
   & 10 (fixed) & $6.51^{6.60}_{6.40}$ & $10.3^{13.6}_{8.96}$ & 10 (fixed)
   & $7.11^{7.59}_{6.83}$  \\
$\tau$
   & $4.02^{4.11}_{3.89}$ & $19.4^{21.1}_{18.3}$ & $3.68^{4.17}_{2.98}$
   & $3.64^{3.84}_{3.48}$ & $4.92^{5.04}_{4.53}$  \\
$A_{\rm C}$ 
   & $1.14^{1.26}_{1.07}$ & $0.724^{0.730}_{0.715}$
   & $2.10^{2.33}_{1.76}$ & $1.58^{1.82}_{1.30}$ & $2.83^{3.08}_{2.64}$  \\
$E_{\rm Fe}$ (keV)
   & $6.447^{6.507}_{6.382}$ & $6.441^{6.468}_{6.409}$
   & $6.399^{6.475}_{6.238}$ & $6.442^{6.513}_{6.350}$ 
   & $6.480^{6.504}_{6.428}$  \\
$\sigma$
   & $0.337^{0.451}_{0.210}$ & $0.210^{0.263}_{0.160}$
   & $0.434^{0.628}_{0.296}$ & $0.319^{0.464}_{0.172}$
   & $0.315^{0.405}_{0.249}$  \\
$A_{\rm Fe}$ 
   & $2.69^{3.28}_{2.29}$ & $2.53^{2.64}_{2.44}$
   & $4.61^{7.31}_{3.58}$ & $3.57^{4.82}_{2.85}$ & $4.40^{5.19}_{3.98}$ \\
EW (eV)
   & $434^{522}_{349}$ & $1820^{1920}_{1750}$ & $381^{550}_{318}$
   & $407^{525}_{337}$ & $396^{443}_{349}$ \\
$L_{\rm X}$ 
   & $2.17 \pm 0.41$ & $1.51 \pm 0.23$ & $3.6 \pm 1.1$ & $2.52 \pm 0.30$ 
   & $2.70 \pm 0.18$ \\
  \tableline
$\chi_{\nu}^{2}$ 
   & 1.20 & 2.60 & 0.899 & 1.03 & 1.68 \\
\enddata
\end{deluxetable}



\end{document}